\documentclass[pra,twocolumn,amsmath,amssymb,floatfix,reprint]{revtex4-1}

\usepackage{graphicx}
\usepackage{dcolumn}
\usepackage{bm}
\usepackage[usenames,dvipsnames]{xcolor}
\usepackage{mciteplus}
\usepackage{braket}

\begin{document}

\title{Dual cascade and dissipation mechanisms in helical
  quantum turbulence}
\author{Patricio Clark di Leoni$^1$\email{clark@df.uba.ar},
  Pablo D.~Mininni$^1$\email{mininni@df.uba.ar}, \&
  Marc E.~Brachet$^2$\email{brachet@physique.ens.fr}}
\affiliation{$^1$Departamento de F\'\i sica, Facultad de Ciencias
Exactas y Naturales, Universidad de Buenos Aires and IFIBA,
CONICET, Ciudad Universitaria, 1428 Buenos Aires, Argentina.\\
  $^2$ Laboratoire de Physique Statistique, \'{E}cole Normale
  Sup{\'e}rieure, PSL Research University; UPMC Univ Paris 06,
  Sorbonne Universit\'{e}s; Universit\'{e} Paris Diderot, Sorbonne
  Paris-Cit\'{e}; CNRS; 24 Rue Lhomond, 75005 Paris, France.}
\date{\today}

\begin{abstract}
    While in classical turbulence helicity depletes nonlinearity and can
    alter the evolution of turbulent flows, in quantum turbulence its
    role is not fully understood. We present numerical simulations of
    the free decay of a helical quantum turbulent flow using the
    Gross-Pitaevskii equation at high spatial resolution. The evolution
    has remarkable similarities with classical flows, which go as far as
    displaying a  dual transfer of incompressible kinetic energy and
    helicity to small scales. Spatio-temporal analysis indicates that
    both quantities are dissipated at small scales through non-linear
    excitation of Kelvin waves and the subsequent emission of phonons.
    At the onset of the decay, the resulting turbulent flow displays
    polarized large scale structures and unpolarized patches of
    quiescense reminiscent of those observed in simulations of classical
    turbulence at very large Reynolds numbers.
\end{abstract}
\maketitle

\section{Introduction}

From the oceans to the solar wind, turbulence is widely found in nature.
It is also observed in quantum fluids such as superfluids and
Bose-Einstein condensates (BECs) \cite{Barenghi14}. Unlike classical
flows, quantum flows have no viscosity and vorticity is concentrated
along topological line defects with quantized circulation
\cite{Feynman55,Donnelly}. While similarities between these two types of
turbulence exist (e.g., both display Kolmogorov spectrum
\cite{Maurer98,Salort12}), there are also
differences \cite{Paoletti08,White10}.

In classical turbulence helicity is an ideal invariant which measures
how tangled vorticity field lines are \cite{Moffatt69}. It is known to
deplete nonlinearities and energy transfer \citep{Kraichnan73},
slow down the onset of dissipation in decaying turbulence and affect 
its dissipation scale \citep{Andre77}, play a role in convective
storms \citep{Lilly86}, and display a dual direct cascade with the
energy \cite{Brissaud73,Moffatt92b}. Its role in quantum turbulence is
less clear. Efforts focus on determining if it is conserved by
studying simple configurations of reconnecting vortex knots
\cite{Scheeler14,Zuccher15,Kleckner16,Clark16,Hanninen16}. Numerical
evidence indicates that in this case helicity is transferred from large
to small scales \cite{Scheeler14,Kleckner16}, and that reconnection or
the transfer of helicity can excite non-linear interacting Kelvin waves
\cite{Fonda14,Clark16}, which eventually may lead to a loss of helicity
by sound emission. Research into the role of helicity in more complex
quantum flows has been lacking, partly due to the difficulties of
quantifying helicity in fully developed turbulent flows. However,
new developments in 3D vortex tracking in Helium experiments
\cite{Lathrop} and in knot generation in BECs \cite{Hall16} provide hopeful
opportunities to tackle this problem.

We present massive numerical simulations of helical quantum turbulence
using the Gross-Pitaevskii equation (GPE) as a model. A quantum version
of the classical Arnold-Beltrami-Childress (ABC) flow is introduced and
used as initial condition to create a helical flow. We use different
methods to quantify helicity, including the regularized helicity
\cite{Clark16} which was shown to give results equivalent to the
centerline helicity for simple knots, and to the classical helicity for
flows with scale separation. We show that helicity is depleted as the
incompressible kinetic energy. As in the classical case
\cite{Brissaud73}, both the incompressible energy and the helicity
follow a Kolmogorov spectrum down to the intervortex distance. At
smaller scales a bottleneck in the energy spectrum is followed by
another Kolmogorov spectrum associated to a Kelvin waves cascade. Energy
and helicity dissipation at coherent length scales is related to Kelvin
waves damping by phonon emision. In physical space, the flow displays
polarized large scale structures formed by a myriad of small scale
knots, and unpolarized quiescent patches mimicking what is observed in
isotropic and homogeneous classical flows at large Reynolds. These
results open up new questions about helicity in quantum flows. In
particular, while successful theories for the energy spectrum exist
\citep{Lvov10}, this is not yet the case for the helicity spectrum.

\section{The Gross-Pitaevskii equation}


The GPE describes the evolution of a zero-temperature condensate of
weakly interacting bosons of mass $m$,
\begin{equation}
    i \hbar \partial_t \Psi = - \hbar^2 (2m)^{-1} \nabla^2 \Psi 
    + g \vert \Psi \vert^2 \Psi ,
    \label{eq:gpe}
\end{equation}
where $g$ is related to the scattering length. Madelung
transformation $\Psi=\sqrt{{\rho}/{m}}\exp{(i m \phi/\hbar)}$ 
relates the wavefunction $\Psi$ to a condensate of density $\rho$ and
velocity ${\bf v}={\bf \nabla} \phi$. Linearizing Eq.~\eqref{eq:gpe}
around a constant $\Psi= \hat{\Psi}_{\bf 0}$ yields the Bogoliubov
dispersion relation 
$\omega_B (k) = c k (1 + \xi^2 k^2/2)^{1/2}$ for sound
waves (or phonons) of speed $c=(g|\hat{\Psi}_{\bf 0}|^2/m)^{1/2}$,
with dispersion taking place at lengths smaller than the 
coherence length $\xi=[g\hbar^2|\hat{\Psi}_{\bf 0}|^2/(2m)]^{1/2}$. 
The Onsager-Feynman quantum of velocity circulation around the
$\Psi=0$ topological defect lines is $h/m$, and the vortex core size
is of order $\xi$ \cite{Proukakis08}.

The GPE conserves the total energy $E$, which can be decomposed as
\cite{Nore97a,Nore97b}: 
\begin{equation}
    E=E_{\rm kin}+E_{\rm int}+E_{\rm q},
\end{equation}
with the kinetic energy $E_{\rm kin} = \langle |\sqrt \rho {\bf
v}|^2/2\rangle$, the internal energy $E_{\rm int}=\langle c^2(\rho
-1)^2/2 \rangle$ and  the so-called quantum energy $E_{\rm q}= \langle
c^2 \xi^2 |\nabla \sqrt{\rho}|^2\rangle$.  The kinetic energy $E_{\rm
kin}$ can be also decomposed into compressible $E_{\rm  kin}^{\rm c}$
and incompressible $E_{\rm kin}^{\rm i}$ components, using $(\sqrt \rho
{\bf v})=(\sqrt \rho {\bf v})^{\rm c}+ (\sqrt \rho {\bf v})^{\rm i}$
with $\nabla \cdot (\sqrt \rho {\bf v})^{\rm i}=0$.

\subsection{Helicity in quantum flows}

The definition of helicity in a classical flow is
\begin{equation}
    H = \int {\bf v}\cdot
    {\bm \omega} \, dV,
\end{equation}
where ${\bm \omega}=\nabla \times {\bf v}$ is the
vorticity.  It follows from Madelung transformation that
\begin{equation}
    {\bm \omega} ({\bf
    r}) = \frac{h}{m} \int d s \frac{d {\bf r}_0}{d s} \delta({\bf r} -
    {\bf r}_0(s)),
\end{equation}
where ${\bf r}_0(s)$ denotes the position of the vortex
lines and $s$ the arclength. Thus vorticity in a quantum flow is a
distribution concentrated along $\Psi=0$ topological line defects where
${\bf v}$ is ill-behaved. In spite of this, some authors
\cite{Zuccher15} compute $H$ by filtering fields to the largest scales
or relying on the regularization introduced by the numerics. Other
authors compute the ``centerline helicity'' by calculating the writhe
and link, two topological quantities which quantify how knotted vortex
lines are, but which require detailed extraction of all centerlines of
the quantized vortices in the flow
\cite{Scheeler14,Laing15,Kleckner16,Krstulovic16}.  Some authors suggest
to also add the twist of equal-phase surfaces (or else just the torsion)
to this definition, but then the total helicity vanishes identically (or
else smoothly formed inflection points change the helicity
discontinuously) \cite{Hanninen16}. A new method which yields the same
results as the ``centerline helicity'' was introduced in \cite{Clark16}
by using the fact that the velocity parallel to the quantized vortex has
only an apparent singularity. The regular smooth velocity oriented along
the vortex line is defined as ${\bf v}_{\rm reg}={v_\parallel {\bf
w}}/{\sqrt{{w}_j{w}_j}}$, where
\begin{equation}
v_\parallel=\frac {\hbar \,
{w}_j\left[(\partial_j \partial_l \Psi) \partial_l
\overline{\Psi}-(\partial_j \partial_l \overline{\Psi}) \partial_l \Psi
\right]} {2 i m \sqrt{{w}_l{w}_l}(\partial_m\Psi)(\partial_m
\overline{\Psi})},
\end{equation}
and 
\begin{equation}
{\bf w} = \frac {\hbar} {im} \bm{\nabla}
\overline{\Psi} \times \bm{\nabla}  \Psi
\end{equation}
(see Appendix~\ref{app} for a detailed
derivation).  The regularized helicity thus reads 
\begin{equation}
H_r= \int {\bf
v}_{\rm reg} \cdot {\bm \omega} \, dV,
\end{equation}
and is well defined in the sense
of distributions \cite{Lighthill58}, as the test function ${\bf v}_{\rm
reg}$ is smooth.  This expression was proven useful even in flows with
hundreds of thousands of knots.

\begin{figure}
    \centering
    \includegraphics[width=8.7cm]{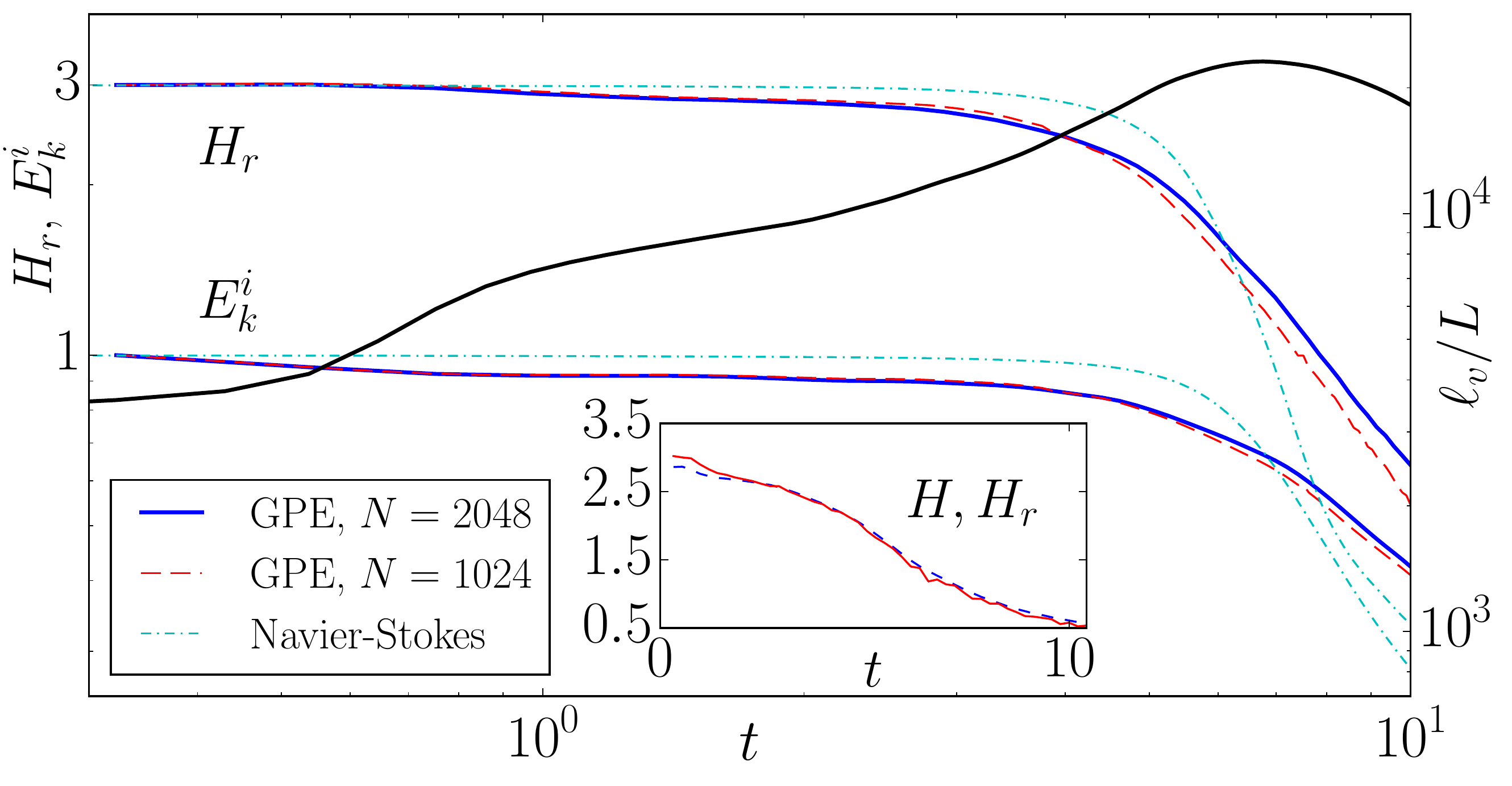}
    \caption{{\it (Color online)} Evolution of the incompressible
      energy $E^i_k$ and of the regularized helicity $H_r$ in the
      $1024^3$ and $2048^3$ GPE runs, and in Navier-Stokes. Note the
      early ``inviscid'' phase in which quantities are approximately
      constant. The solid black line shows the total vortex length in
      the $2048^3$ GPE run. {\it Inset:} $H_r(t)$ (dashed blue line) and
      the non regularized helicity $H(t)$ (solid red line) in the
      $2048^3$ GPE run.} 
    \label{timeevol}
\end{figure}

\subsection{Numerical simulations}

The GPE is solved in its dimensionless form and all quantities presented
here are dimnesionless (see \cite{Nore97b,Clark15a} for more details).
All numerical simulations in this paper have mean density $\rho_0=1$.
Physical constants in Eq.~\eqref{eq:gpe} are determined by $\xi$ and
$c=2$, and the quantum of circulation $h/m = c \xi/\sqrt{2}$.
Simulations were performed using $512^3$, $1024^3$, and $2048^3$ grid
points, in a domain of linear size $L=2\pi$. The largest $2048^3$ GPE
simulation has a healing length $\xi \approx 2.2 \times 10^{-3}$ and an
intervortex distance $\ell \approx 8\times 10^{-2}$ (computed as in
\cite{Nore97a,Nore97b}). As a comparison, in $^3He$ experiments the size
of the vortex core is $\approx 10^{-8}$ m, the intervortex distance is
$\approx 10^{-5}$ m, and the system size is of order $10^{-2}$ m
\cite{Barenghi14}. Scale separation is smaller for BECs, which are also
better modeled by the GPE. Although proper scale separation in a
simulation is currently out of reach, the $2048^3$ run is a significant
improvement over most simulations of quantum turbulence which have one
order or magnitude difference between $L$ and $\xi$.

\begin{figure}
    \centering
    \includegraphics[width=8.7cm]{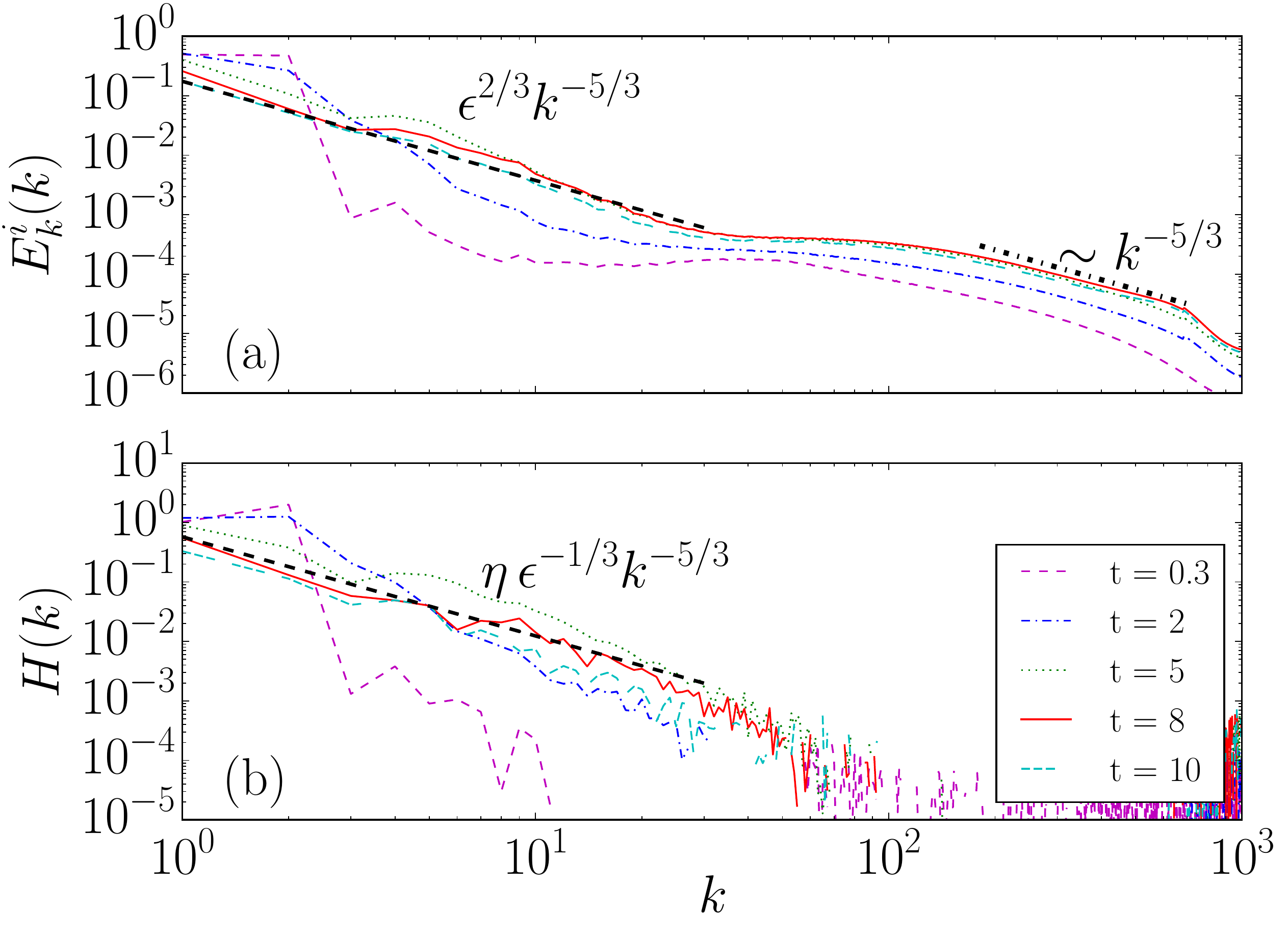}
    \caption{{\it (Color online)} Spectrum of the (a) incompressible
    kinetic energy, and (b) helicity in the $2048^3$ GPE run. At large
    scales both follow a scaling compatible with a classical dual
    cascade (thick dashed lines). At scales smaller than the intervortex
    scale ($k_\ell \approx 80$) a second range compatible with a Kelvin
    wave cascade is observed in $E^i_k$ (thick dash-dotted line). The
    helicity spectrum broadens in time indicating a direct transfer.}
    \label{kspectrum}
\end{figure}

\begin{figure}
    \centering
    \includegraphics[width=8.7cm]{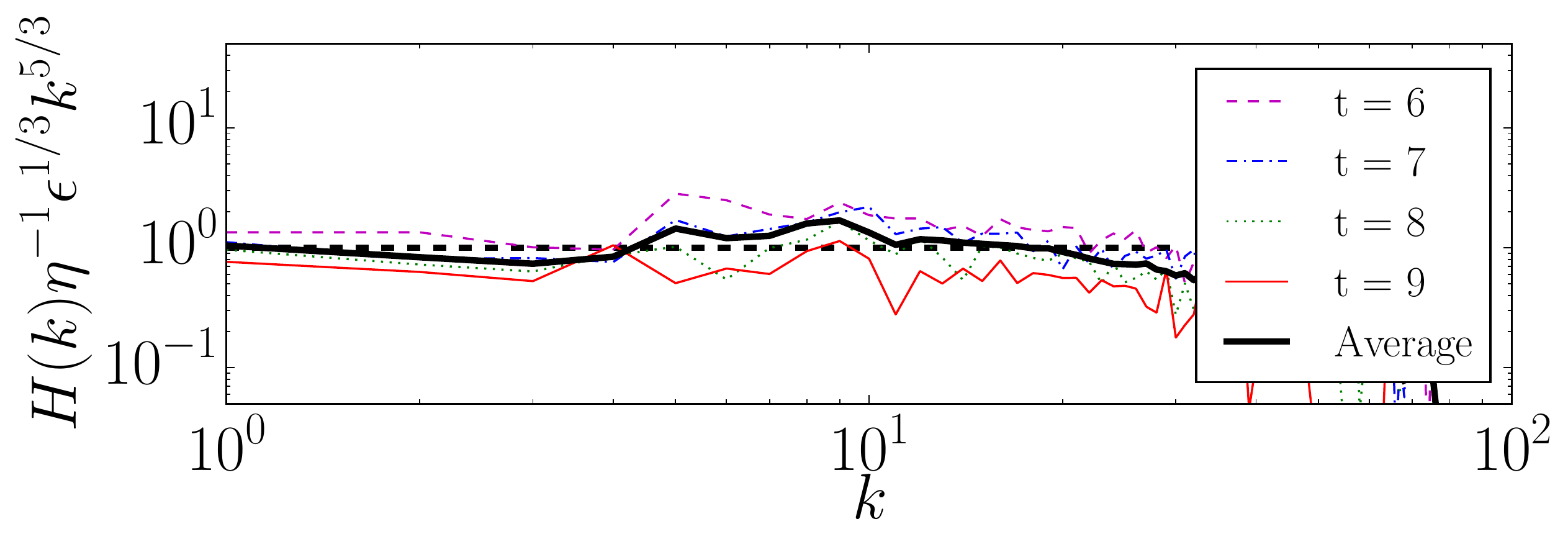}
    \caption{{\it (Color online)} Compensated helicity spectra in the
        $2048^3$ GPE run. The spectrum is compatible with 
        $\eta \epsilon^{-1/3} k^{-5/3}$ scaling. The time-averaged
        spectrum is also shown.} 
    \label{compensated}
\end{figure}

To compare the GPE simulations with classical ABC flows we
also simulated the incompressible Navier-Stokes (NS) equation 
\begin{equation}
\partial _t \bf{u}+({\bf u} \cdot \nabla) {\bf u} =- \nabla p +\nu
    \nabla^2 {\bf u},
\end{equation}
with $\nabla \cdot \bf{u}=0$, using $512^3$ points and
viscosity $\nu=6.5\times10^{-4}$. All equations were integrated 
 using GHOST \cite{Mininni11}, a 
pseudospectral code with periodic boundary conditions, 
fourth order Runge-Kutta to compute time derivatives, and the $2/3$
rule for dealiasing.

As initial condition we use a superposition of $k=1$ and $k=2$ basic
ABC flows: 
${\bf v}_{\rm ABC}={\bf v}_{\rm ABC}^{(1)}+{\bf v}_{\rm ABC}^{(2)}$,
with
\begin{eqnarray}
{\bf v}_{\rm ABC}^{(k)} = & \left[ B \cos(k y) + C \sin(k z) \right] \hat{x} 
+ \left[ C \cos(k z) + \right. \nonumber \\
  {}& \left. A \sin(k x)  \right] \hat{y} +
    \left[ A \cos(k x) + B \sin(k y) \right] \hat{z},
\label{ABC}
\end{eqnarray}
and $(A,B,C)=(0.9,1,1.1)/\sqrt{3}$.
The basic ABC flow is a $2\pi$-periodic stationary solution of the
Euler equation with maximal helicity. To build its quantum version 
 we first take the flow with $B=C=0$ and use Madelung
transformation to obtain a wavefunction
$\Psi_{A,k}^{x,y,z} =\exp\{i [A \sin(k x)\,m/\hbar] y
    +i [A \cos(k x)\,m/\hbar] z\}$, 
where $[a]$ stands for the nearest integer to $a$ to enforce
periodicity. The wavefunction of the quantum ABC flow is then obtained
as 
$\Psi_{\rm ABC}^{(k)}= \Psi_{A,k}^{x,y,z}  \times \Psi_{B,k}^{y,z,x}  
    \times \Psi_{C,k}^{z,x,y} $.
Finally, 
$\Psi_{\rm ABC}=\Psi_{\rm ABC}^{(1)} \times \Psi_{\rm ABC}^{(2)}$ 
corresponds to the initial flow ${\bf v}_{\rm ABC}$. In practice, to
correctly set the initial density with defects along the
vortex lines and to correct frustration errors arising from
periodicity, following \cite{Nore97a,Nore97b} we first evolve
$\Psi_{\rm ABC}$  using the advected real
Guinzburg-Landau equation 
\footnote{$\partial_t \Psi=  \frac{\hbar}{2 m} \nabla^2 \Psi 
    +(\frac{g\rho_0}{m}-g|\Psi |^2
    -\frac{m {\bf v}_{\rm ABC}^2}{2 \hbar})\Psi
    -i {\bf v}_{\rm ABC} \cdot \nabla  \Psi$}, 
whose stationary solutions are solutions of the GPE with minimal
emission of acoustic energy.

\begin{figure}
    \centering
    \includegraphics[width=8.7cm]{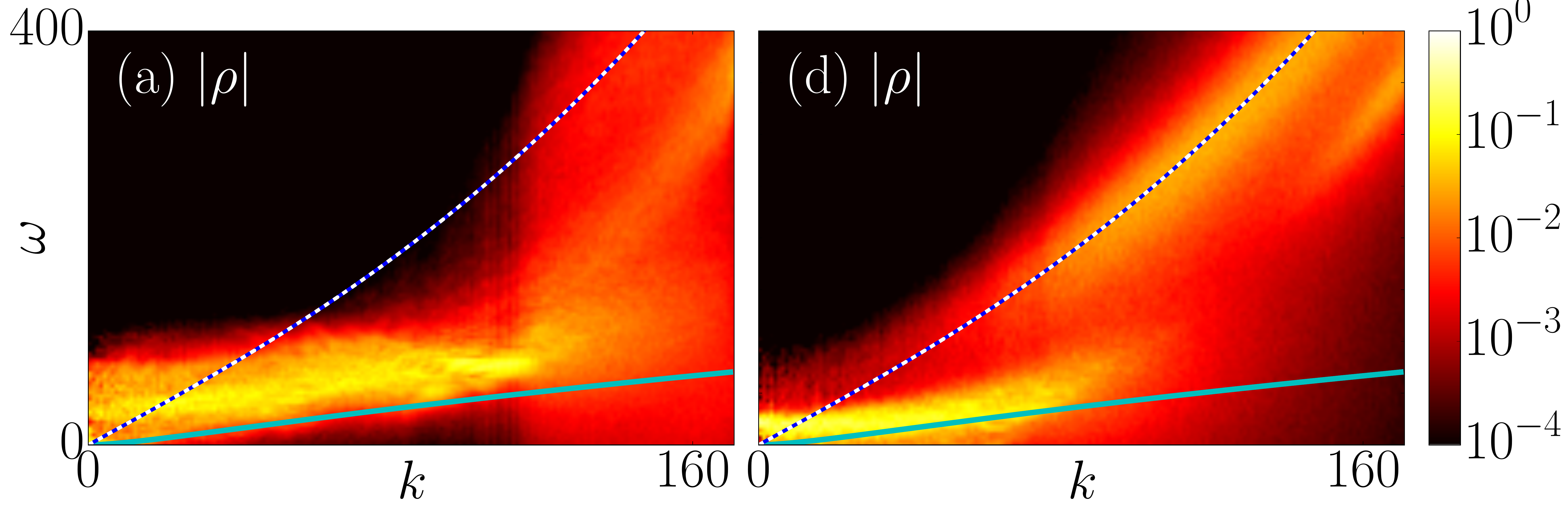}
    \includegraphics[width=8.7cm]{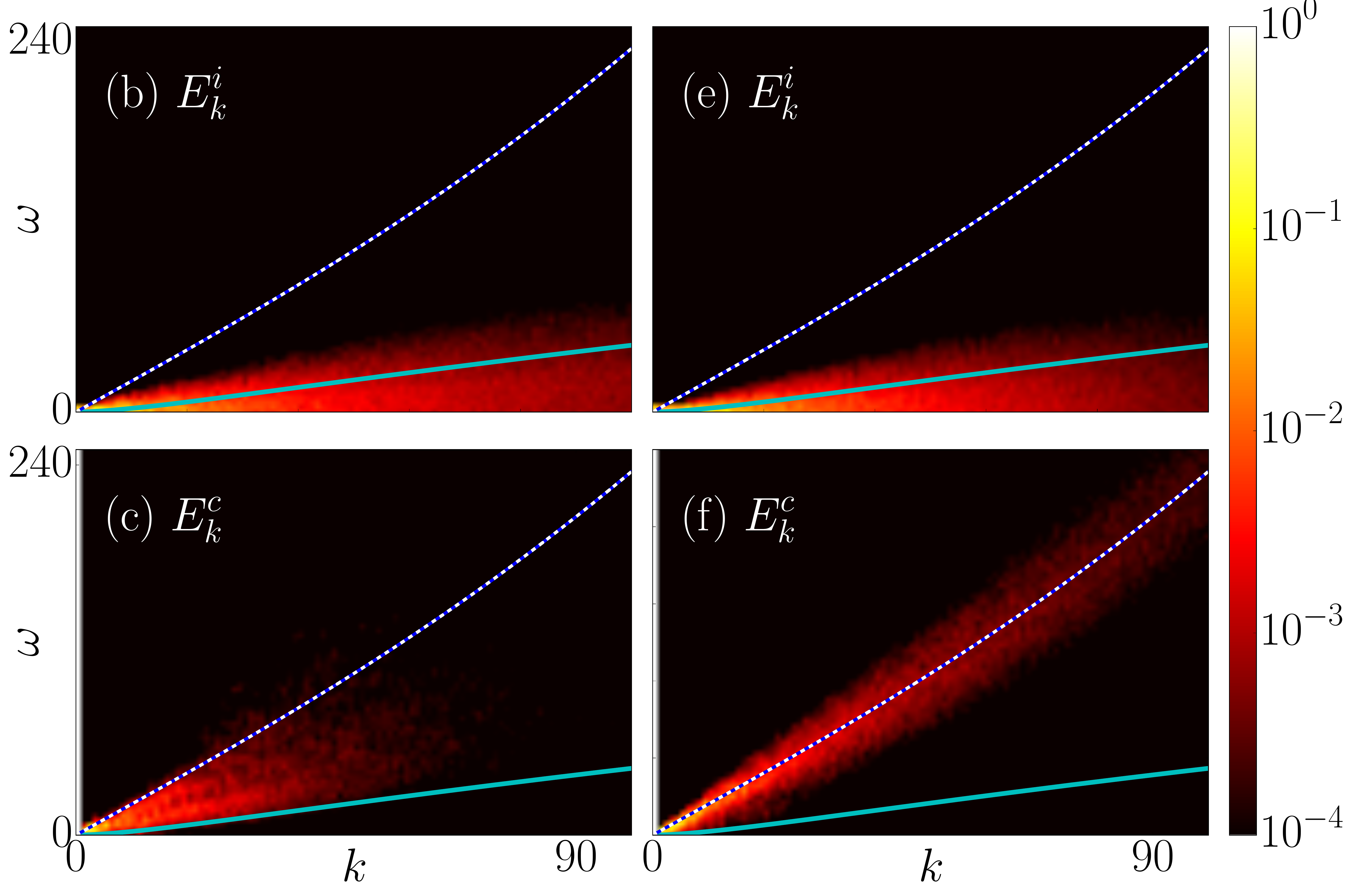}
    \caption{{\it (Color online)} Spatio-temporal power spectra for
      the $512^3$ GPE run between $t=0$ and $2$ for (a) the mass
      density $\rho$, and zooms between $k=0$ and $100$ for 
      (b) the incompressible and (c) compressible velocity. Same for
      late times ($t \in [8,10]$) are shown in (d), (e), and (f). The
      solid (green) curve is the dispersion relation of Kelvin waves,
      while the dotted (white) curve corresponds to sound waves.}
    \label{spectra}
\end{figure}

\section{Results} 

In Fig.~\ref{timeevol} we show the evolution of the incompressible
kinetic energy $E^i_k$ and of the regularized helicity $H_r$ for the
$1024^3$ and $2048^3$ GPE runs, and for the 
NS equation (with $H_r\equiv H$). In all cases, $E^i_k$ and
$H_r$ remain approximately constant for the first turnover times while
turbulence develops (the so-called ``inviscid'' phase in the decay of
classical flows). The total vortex length $\ell_v$ peaks at the end of
this phase, which ends slightly earlier for $H_r$ than for
$E^i_k$. Afterwards, $H_r$ and $E^i_k$ decrease  in
what seems a self-similar decay, with different rates in each
system. As  total energy is conserved in GPE,
the decay of $E^i_k$ is accompanied by a growth of the other
components of the energy, in particular of $E^c_k$. Indeed, in quantum
turbulence the decay of $E^i_k$ is expected to result from the
emission of phonons \cite{Vinen02}, and thus from classical results
\cite{Teitelbaum11} the decay in $E^i_k$ should also produce a decay
in $H_r$.

\begin{figure}
    \centering
    \includegraphics[height=9cm]{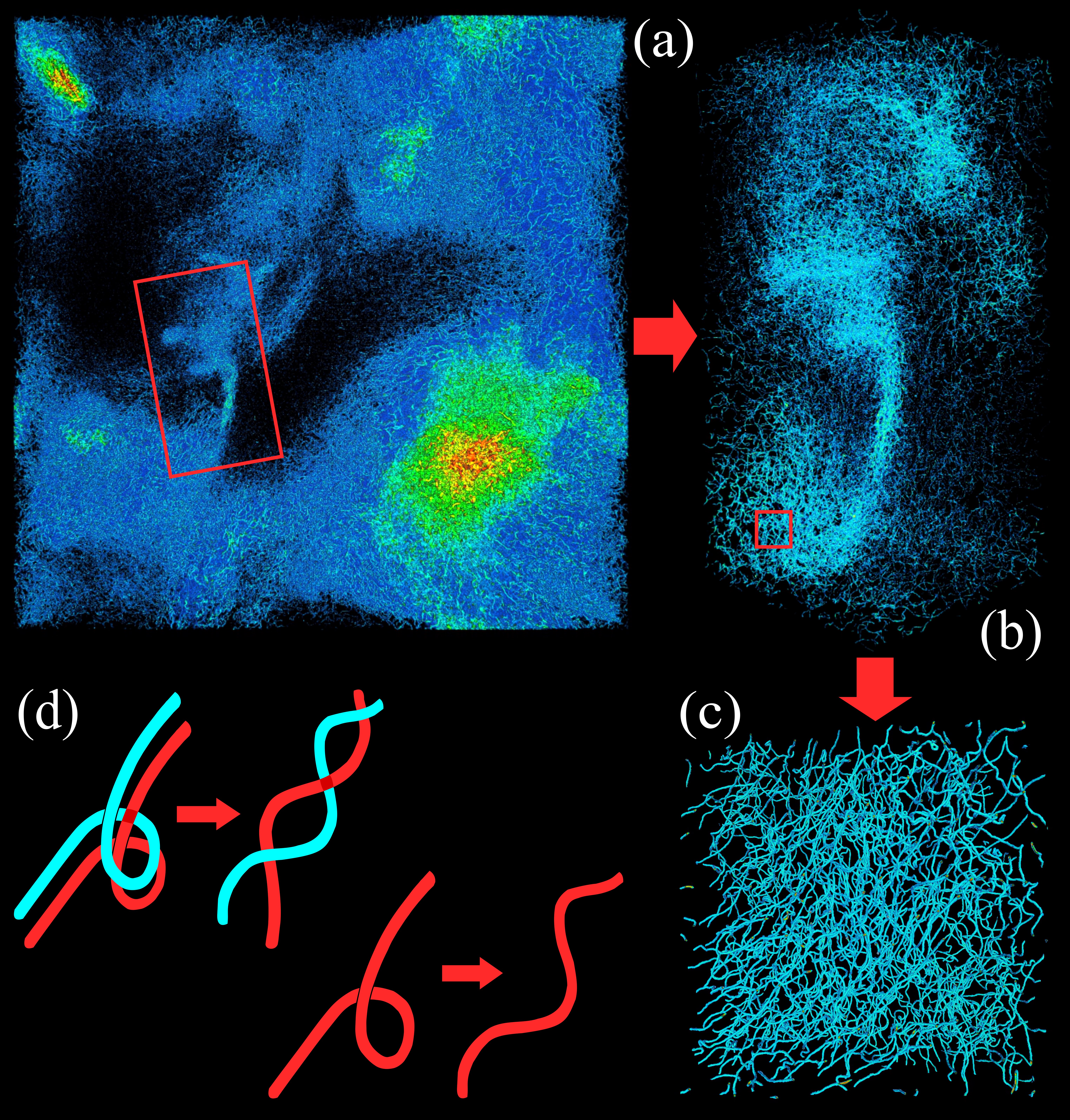}
    \caption{{\it (Color online)} Three-dimensional rendering of 
      vortex lines at the onset of the decay in the $2048^4$ GPE
      run of (a) a slice of the full box, and succesive zooms 
      in (b) and (c) into the regions indicated by the (red)
      rectangles. (d) Sketch of the transfer of helicity from writhe
      to twist in a bundle of vortices, and for an individual vortex.}
    \label{vapor}
\end{figure}

The inset in Fig.~\ref{timeevol} compares  $H$ (non regularized)
and  $H_r$ in the $2048^3$ GPE run. Both are in
good agreement, but $H_r$ is smoother and less noisy, making it a better
fit to study the global evolution of helicity in quantum turbulence.
However, the agreement between $H_r$ and $H$ allows us to use $H$ to
compute spectra.


Figure \ref{kspectrum} shows spectra of $E^i_k$ and $H$ at different
times in the $2048^3$ GPE run. The spectra build up rapidly from the
initial conditions, and the energy and helicity excite larger
wavenumbers as time increases. At $t=5$ both already display inertial
ranges. At large scales they follow a power law compatible
with a classical dual energy and helicity cascade \cite{Brissaud73},
with Kolmogorov constant $C_K\approx 1$
\begin{equation}
E(k) \approx \epsilon^{2/3}k^{-5/3},\hspace{0.5cm} H(k) \approx
\eta\epsilon^{-1/3}k^{-5/3}\label{eq:speciner} ,
\end{equation}
and with $\epsilon$ and $\eta$ calculated directly using
\begin{equation}
\epsilon = - dE^i_k/dt , \,\,\,\, \eta = -dH/dt ,
\end{equation}
from the data in Fig.~\ref{timeevol} after the onset of decay. Around
the mean intervortex scale ($k_\ell \approx 80$) $E^i_k$ diplays a
bottleneck compatible with predictions in \cite{Lvov10}. This
bottleneck is followed by an inertial range $\sim k^{-5/3}$
 predicted for a Kelvin wave cascade
\cite{Lvov10,Krstulovic12} and which below is confirmed for the lower
resolution run. Figure \ref{compensated} shows compensated
  helicity spectra, which corroborates the behavior observed in
  Fig.~\ref{kspectrum}.

Of particular interest is the evolution of $H(k)$. For early times $H$
is concentrated at low wavenumbers, as expected for the initial
condition. But later it is transferred to larger wavenumbers as the
cascade-like spectrum develops. While there is no rigourous proof of
conservation of helicity in  quantum flows, note that using the Hasimoto
transformation \cite{Hasimoto72} the evolution of a vortex line can be
mapped into a nonlinear Schr\"odinger equation which conserves momentum.
But momentum of a vortex line (e.g., the translation of the centerline
in the direction of vorticity) can result in net helicity. Thus, 
vortex lines evolution could indeed conserve
helicity (except for depletion by emission of phonons). At small
scales $H(k)$ displays wild fluctuations (in amplitude and sign), which
is to be expected as the non-reguralized
helicity is ill-behaved at those scales. The fact that the helicity
dynamics, at least at large scales, of a quantum flow mimics those of a
classical one is remarkable. But it also begs the question of what
happens to the helicity at scales smaller than the intervortex distance.
Indications exist that Kelvin waves carry helicity
\cite{Scheeler14,Clark16}, but such a possibility requires confirmation
of their presence.

\begin{figure}
    \centering
    \includegraphics[width=8.7cm]{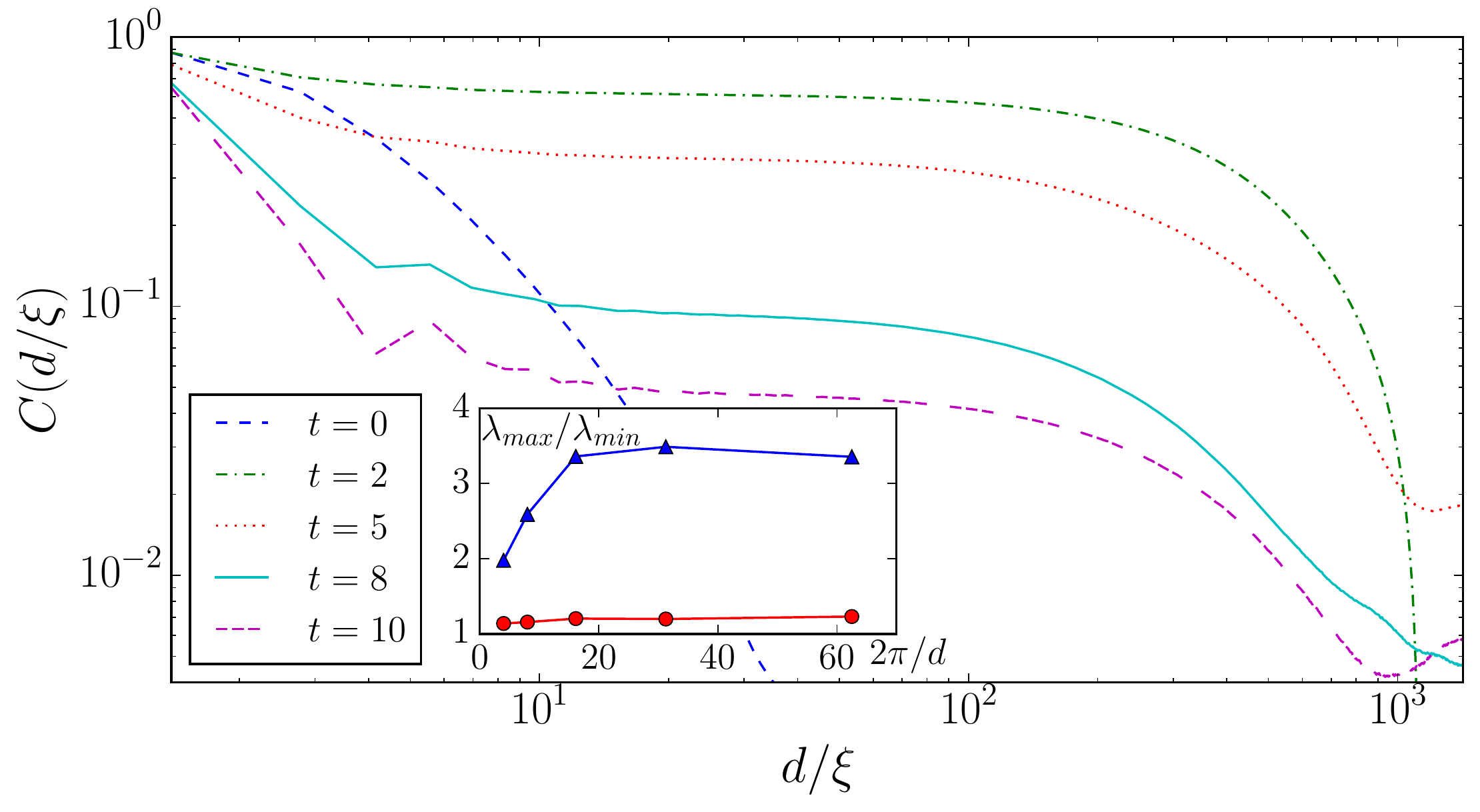}
    \caption{{\it (Color online)} Correlation function of $\rho$ in
      the $2048^3$ GPE run. At $t=0$ it decays rapidly in units of the
      healing length $\xi$, but then quickly develops long-range
      correlations. Inset: Ratio of eigenvalues
      $\lambda_{max}/\lambda_{min}$ as a function of $2 \pi/d$, with
      $d$ the size of the box used for the average (blue triangles:
      regions with structures, red triangles: regions of quiescence).}
    \label{corr}
\end{figure}

To verify this, as well as phonons being the dissipation mechanism for
$E^i_k$ and $H$, we must detect Kelvin and
sound waves. To do this we use the spatio-temporal spectrum
\cite{Clark15a}, i.e., the four-dimensional power spectrum of a 
field as a function of wave vector and frequency. The spectrum allows
quantification of how much power is in each mode $({\bf k}, \omega)$,
and waves can be separated from the rest as they satisfy a known
dispersion relation $\omega({\bf k})$. As its computation requires
huge amounts of data (i.e., storage of fields resolved in space and
time), we compute it for the $512^3$ GPE run. Figure \ref{spectra}
shows this spectrum (after integration in ${\bf k}$ using isotropy to
obtain  dependency on $k$ and $\omega$) for $\rho$, and zooms for small
$k$ for $E^i_k$ and $E^c_k$, for early and late times (respectively,
$t\in [0,2]$ and $t\in [8,10]$). The dispersion relation of Kelvin
$\omega_K (k)$
\footnote{$\omega_K (k) = \{1 \pm [1 + k a
      {K_0(ka)}/{K_1(ka)}]^{1/2}\}\sqrt{2} c \xi /a^2$, 
  where $a$ is the vortex core radius and $K_0$, $K_1$ are
  modified Bessel functions. This dispersion relation is quadratic for
  small $k$ and linear for large $k$.}
and sound waves $\omega_B (k)$ are shown. Note that,
compared with the $2048^3$ run, $\xi$ in this run is 4 times
larger, and values of $k$ are 4 times smaller.

At early times, power in the spatio-temporal spectrum of $\rho$ is
broadly spread over modes that do not correspond to waves. $E^i_k$
shows some excitations compatible with Kelvin waves, and $E^c_k$ has
little energy with no phonon excitation. At late times power in
fluctuations of $\rho$ moves towards the Kelvin wave
dispersion relation up to $k\approx 80$, and then the power jumps 
towards the dispersion relation of phonons. The spectra $E^i_k$
and $E^c_k$ confirm this picture, with power concentrating in
$E^c_k$ in the vicinity of the sound dispersion relation. Exploration
of these spectra for different time ranges shows that as time evolves
more energy goes from Kelvin wave modes to phonons. While this
analysis is performed at lower resolution and thus wavenumbers 
for the transition are smaller than in the $2048^3$ run,
the spectra confirm the dynamics  in Figs.~\ref{timeevol} and
\ref{spectra}: with time, energy and helicity go from large to smaller
scales in which Kelvin waves are excited, and they are
finally dissipated into phonons.


This can be further confirmed by visualizing vortices in real space
at the onset of decay. Figure~\ref{vapor} shows a
three-dimensional rendering of quantum vortices at $t\approx 2.5$ 
in the $2048^3$ GPE run. Large-scale eddies, formed up by a myriad of
small-scale and knotted vortices, emerge. Similar behavior
has been observed at finite temperature quantum turbulence 
simulations, where the bundle was correlated with high vorticity
in the normal fluid component \cite{Morris08,Nemirovskii13}. At zero temperature two
results \cite{Sasa11,Baggaley12,Baggaley12b} also hinted at this 
behavior, but in none the fine structure of the vortex bundle was
resolved. More importantly, the large scale flow shows inhomogeneous
regions with high density of vortices and quiet regions with low
density. These large-scale patches were not present in the initial
conditions (which have homogeneously distributed vortices) and are
created by the evolution as shown below. The patches are reminiscent
of those observed in isotropic and homogeneous turbulence at high
resolution in non-helical \cite{Ishihara09} and helical
\cite{Mininni06} flows, further confirming the similarity between 
quantum and classical turbulence at scales larger than the intervortex
separation.


The spontaneous emergence of large-scale correlations in the system
can be confirmed by the spatial correlation function 
\begin{equation}
C(d) = \left< (\rho({\bf x} +d\hat{x}) - \rho_0)(\rho({\bf x}) -
  \rho_0) \right>, 
\end{equation}
shown in Fig.~\ref{corr}. This correlation function is related to the
internal energy spectrum by the Wiener-Khinchin theorem. At $t=0$,
$C(d)$ decays rapidly in units of the healing length $\xi$,  and it is
dominated by the vortex core size. But the system rapidly develops
long-range correlations (up to $\approx 1000 \xi$) and later $C$ decays
in a self-similar way.  Furthermore, computing the ratio of eigenvalues
$\tau = \lambda_{max}/\lambda_{min}$ for the tensor $\langle
\partial_i\rho \partial_j\rho \rangle$ averaged in boxes of size $1/10$
of the linear domain size, typically yields $\tau\approx 3$ in regions
with large scale structures and $\tau \approx 1$ in quiescent regions,
indicating anisotropy and a copious vortex polarization in the former
(see Fig.~\ref{corr} inset). 

\section{Conclusions}

The results indicate that helicity can be conserved in quantum
turbulence at large-scales and as it is transferred towards smaller
scales (see Fig.~\ref{kspectrum}), but eventually it decays through the
emission of phonons produced by a Kelvin wave cascade
(Fig.~\ref{spectra}).  We can draw a comparison with the classical case,
where now a bundle of quantum vortices (as seen in Fig.~\ref{vapor})
would play the role of classical vortex tubes. Tubes, in contrast to
lines, add an extra degree of freedom to the helicity: their
twist. Thus in the classical and quantum cases, large scale helicity
can be transformed from writhe to twist for a bundle of vortices (see
Fig.~\ref{vapor}.d). But for individual quantum vortices, the transfer
(e.g., through reconnection) would result in the excitation of a
Kelvin wave which can eventually be damped. This indicates that
individual quantum vortex lines behave like classical vortex tubes
with a mechanism to relax the twist, and as such, the correct analogy
between classical and quantum flows only holds for scales larger than
$\ell$ for which bundles of quantum vortices behave as classical
vortex tubes.

\appendix
\section{Derivation of the regularized velocity}
\label{app}

To calculate the helicity in a quantum flow we need information of both
the velocity and the vorticity along the vortex lines. This is
problematic as both quantities have singularities along those lines.
Therefore, we need to regularize one of them in order to have a
well-behaved integral for the helicity (in the sense of distributions
\cite{Lighthill58}). Although in principle it may seem possible to
regularize any of the two fields, the choice of regularizing the
velocity and not the vorticity is not arbitrary. In the Gross-Pitaevskii
equation, the vorticity is correctly described by a distribution.
Instead, the only component of the velocity that is not well behaved is
the one perpendicular to the vortex line. But for the calculation of the
helicity we need the parallel component, whose problem is to have a
$0/0$ indeterminacy in its defintion. Thus, regularizing the velocity
allows us to keep its well defined component which contributes to the
helicity, while leaving the vorticity as a Dirac delta distribution also
allows us to not bother with the values of the regularized field outside
the vortex line, which should give no contribution to the helicity. Here
we outline a detailed explanation of how to derive the regularized
velocity, from which the expression of the regularized helicity follows
immediately.

The velocity of the superfluid is given by
\begin{equation}
    {\bf v} = \frac{ \hbar}{2 i m} \frac{\bar{\Psi} \bm{\nabla} \Psi - \Psi
    \bm{\nabla}  \bar{\Psi}}{\Psi \bar{\Psi}} .
\label{veltot}
\end{equation}
Without loss of generality we can suppose that there is a vortex line
going through ${\bf r}=0$ (the radial cylindrical vector) in the
direction of the $z$-axis.  Let us define the unit vector basis $({\bf
  \hat e}_x,{\bf \hat e}_y,{\bf \hat e}_z)$.  The existence of a
vortex line passing through $\bf{r}=0$ 
and pointing in the $z$-direction implies that $\Psi(0)=0$,
$\bar\Psi(0)=0$, ${\bf \hat e}_z \cdot \bm{\nabla} \Psi(0)=0$, 
and ${\bf \hat e}_z \cdot \bm{\nabla} \bar\Psi(0)=0$. Thus
$\bm{\nabla} \Psi(0)$ and $\bm{\nabla} \bar\Psi(0)$ are linear
combinations of ${\bf \hat e}_x$ and ${\bf \hat e}_y$. 
Taylor-expanding to first order the numerator and denominator of the
above expression for ${\bf v}({\bf r})$ around ${\bf r}=0$ one finds
\begin{align}
\Psi(x,y,z) &=x \partial_x \Psi(0)+y \partial_y \Psi(0) +
\mathcal{O}({\bf r}^2) ,
\\
 \bar \Psi(x,y,z) &=x \partial_x \bar \Psi(0)+y \partial_y \bar \Psi(0)
+ \mathcal{O}({\bf r}^2), 
\\
\bm{\nabla}  \Psi(x,y,z) &= \bm{\nabla}  \Psi(0) + {\bf r}\cdot
\bm{\nabla} (\bm{\nabla} \Psi)(0)+\mathcal{O}({\bf r}^2) ,
\\
\bm{\nabla} \bar \Psi(x,y,z) &= \bm{\nabla}  \bar \Psi(0) + {\bf r}\cdot
\bm{\nabla} (\bm{\nabla} \bar \Psi)(0)+\mathcal{O}({\bf r}^2) .
\end{align}

After replacing the above expressions in Eq.~\eqref{veltot} and
dropping quadratic terms, the perpendicular ($x$ and $y$) components
of the velocity diverge in the limit ${\bf r} \to 0$, as 
${\bf v}_\perp$  reads
\begin{equation}
{\bf v}_\perp({\bf r}) = \frac{\hbar}{2 i m} \left(\frac{\nabla \Psi(0)
    }{x \partial_x \Psi(0)+y \partial_y \Psi(0)} - c.c\right).
\end{equation}

On the other hand, the velocity component parallel to the centerline
vorticity $v_\parallel({\bf r})={\bf v}({\bf r})\cdot {\bf \hat e}_z$ reads
\begin{align}
\nonumber
v_\parallel({\bf r})=\frac{\hbar}{2 i m}
     & \left( \frac{x (\partial_{xz} \Psi)(0)
                 + y (\partial_{yz} \Psi)(0)
                 + z (\partial_{zz} \Psi)(0)}
                  {x\partial_x \Psi(0)+y \partial_y \Psi(0)}
              \right.
\\
&\left. -c.c. \vphantom{\frac12}\right),
\label{vp1}
\end{align}
which is finite in the limit ${\bf r} \to 0$. This last expression for
$v_\parallel(\bf r)$ can be seen as resulting from l'H\^opital's rule 
applied to the limit of $v_\parallel{(\bf r})$ when ${\bf r}\to 0$ in
the direction $(x,y,z)$.  The limit obviously depends on the direction
as, in deriving the above formulae, the only hypotheses we have made
are that $\Psi$ is sufficiently differentiable and has a zero-line
directed toward $z$.

In order to turn the above expression into a workable ansatz for
$v_\parallel(0)$, we need to pick a reasonable direction along which
$\Psi$ will have a significant variation. The simplest vectors we
have at point ${\bf r}=0$, perpendicular to the vortex line and 
satisfying the condition, are $\bm{\nabla}  \Psi(0)$ and 
$\bm{\nabla}  \bar \Psi(0)$. Thus we can multiply the first term in the
r.h.s.~of Eq.~\eqref{vp1} by $\bm{\nabla} \bar \Psi$, and its complex
conjugate by $\bm{\nabla}  \Psi$ in order to maintain the reality
of the velocity field. In this way we arrive to the following
expression
\begin{equation}
v_\parallel(0)=\frac{\hbar}{2 m \,i} (\frac {\partial_x \bar \Psi
    \partial_{xz} \Psi + \partial_y \bar \Psi  \partial_{yz}\Psi +
    \partial_z \bar \Psi \partial_{zz}\Psi}{\partial_x \bar \Psi
    \partial_x \Psi+\partial_y \bar \Psi  \partial_y \Psi+\partial_z
    \bar \Psi  \partial_z \Psi}  -c.c.) .
\end{equation}

A first check that this ansatz is reasonable is to plug in  $\Psi \sim
(x+ i y) e^{i z U_z m/\hbar}$ and explicitly verify that this gives
$v_\parallel(0)=U_z$. Further validations were performed in
\cite{Clark16}, where it was shown that the helicity computed with the
regularized velocity agrees with the topological definitions of
writhe, link, and twist. Also, in \cite{Clark16} it was shown that
this expression gives the correct value of helicity for different knots,
and that in quantum flows with helicity it gives a value that matches
the helicity in the equivalent classical large-scale helical flow.

As a final remark, it is important to note that for arbitrarily
aligned vortex lines, the direction parallel to the vortex line
($\hat{z}$ in the particular case considered above) can be easily
obtained by doing the vector product between $\bm{\nabla}  \Psi$ and
$\bm{\nabla}  \bar \Psi$. 

\begin{acknowledgements}
The authors acknowledge financial support from Grant No.~ECOS-Sud
A13E01, and from computing hours in the CURIE supercomputer granted by
Project TGCC-GENCI No.~x20152a7493.
\end{acknowledgements}

\bibliography{ms}

\end{document}